\documentclass[journal,twoside,web]{ieeecolor}
\usepackage{generic}
\usepackage{cite}
\usepackage{amsmath,amssymb,amsfonts}
\usepackage{graphicx}
\usepackage{algorithm}
\usepackage{hyperref}
\hypersetup{hidelinks}
\usepackage{textcomp}

\usepackage{algpseudocode}
\usepackage{multirow}
\usepackage{url}

\usepackage[nolist,nohyperlinks]{acronym}
\begin{acronym}
\acro{DL}{deep learning}
\acro{FC}{functional connectivity}
\acro{GNN}{graph neural networks}
\acro{MRI}{magnetic resonance imaging}
\acro{fMRI}{functional magnetic resonance imaging}
\acro{GAT}{graph attention networks}
\acro{ASD}{Autism Spectrum Disorder}
\acro{ADHD}{Attention-deficit Hyperactivity Disorder}
\acro{AD}{Alzheimer's Disease}
\acro{RE-CONFIRM}{Robust Evaluation of CONnectome Features Identified by Relevance Measures}
\acro{ML}{machine learning}
\acro{XAI}{explainable AI}
\acro{FM}{foundation models}
\end{acronym}

\usepackage{booktabs}

\def\BibTeX{{\rm B\kern-.05em{\sc i\kern-.025em b}\kern-.08em
    T\kern-.1667em\lower.7ex\hbox{E}\kern-.125emX}}
\begin{document}

\title{Foundation models for discovering robust biomarkers of neurological disorders from dynamic functional connectivity}

\author{Deepank Girish*, Yi Hao Chan*, Sukrit Gupta, Jing Xia, and Jagath C. Rajapakse, \IEEEmembership{Fellow, IEEE}
\thanks{This research is supported by AcRF Tier-1 grant RG18/26 of Ministry of Education, Singapore.}
\thanks{Deepank Girish, Yi Hao Chan, and Jagath C. Rajapakse are with the College of Computing and Data Science, Nanyang Technological University, Singapore.}
\thanks{Sukrit Gupta is with the Department of Biomedical Engineering and the School of Artificial Intelligence and Data Engineering, Indian Institute of Technology Ropar, India.} 
\thanks{Jing Xia is with the College of Instrument and Computer Science, Zhejiang University, China.}
\thanks{* These authors contributed equally.}
\thanks{Corresponding author: Jagath C. Rajapakse (e-mail: ASJagath@ntu.edu.sg).}}

\maketitle

\begin{abstract}
Several brain foundation models (FM) have recently been proposed to predict brain disorders by modelling dynamic functional connectivity (FC). While they demonstrate remarkable model performance and zero- or few-shot generalization, the salient features identified as potential biomarkers are yet to be thoroughly evaluated. We propose RE-CONFIRM, a framework for evaluating the robustness of potential biomarker candidates elucidated by deep learning (DL) models including FMs. \textcolor{black}{Experiments across five datasets covering \ac{ASD}, \ac{ADHD}, and \ac{AD}, including two external validation datasets, showed that commonly used performance metrics provide an intuitive assessment of predictive performance but are insufficient to evaluate the robustness of the identified biomarkers.}
RE-CONFIRM metrics revealed that simply finetuning FMs leads to models that fail to capture regional hubs effectively, even in disorders where hubs are known to be implicated, such as ASD and ADHD. \textcolor{black}{In view of this, we propose Hub-LoRA (Low-Rank Adaptation) as a fine-tuning technique that enables FMs to achieve better performance than customised DL models but also produce neurobiologically plausible biomarkers supported by meta-analyses.} RE-CONFIRM is generalizable and can be easily applied to ascertain the robustness of DL models trained on functional MRI datasets. 
Code is available at: \url{https://github.com/SCSE-Biomedical-Computing-Group/RE-CONFIRM}
\end{abstract}


\section{Introduction}
\label{sec:introduction}

\begin{figure*}[ht]
    \centering
    \includegraphics[width=0.875\linewidth]{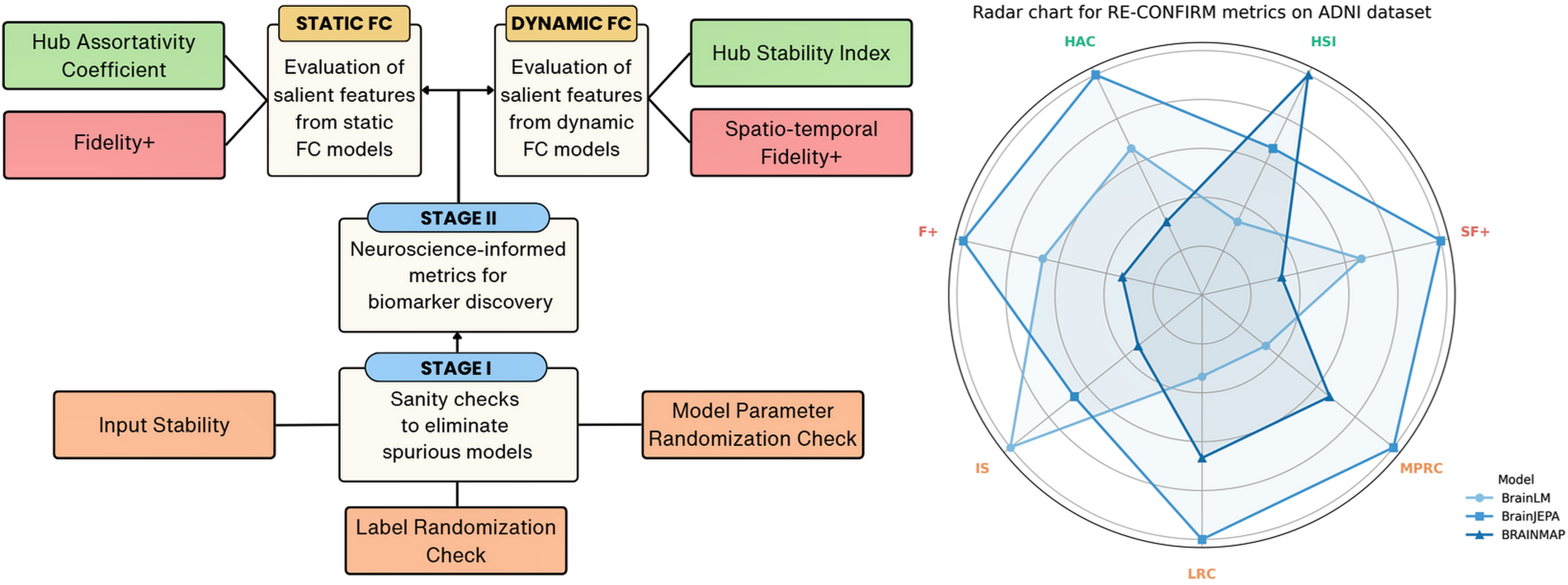}
    \caption{Overview of the proposed RE-CONFIRM framework. It encompasses both technical aspects (stability, label randomization check, etc.) and connectome-specific aspects (e.g. hub assortativity coefficient, hub stability index), for evaluating both static and dynamic functional connectivity biomarkers. When used together with classical metrics such as model accuracy, a comprehensive evaluation of the robustness of model performance can be achieved. Radar plots facilitate comparisons across models by visualising relative rankings for each metric.}
    \label{fig:reconfirm}
\end{figure*}

\IEEEPARstart{O}{ver} the past decade, \ac{DL} models such as \ac{GNN} \cite{chan2025discovering} and transformer-based architectures \cite{cong2024comprehensive} have been increasingly used on \ac{fMRI} scans to discover potential biomarkers of brain disorders. This is often achieved by using intrinsically interpretable models (e.g., models with the attention mechanism) or applying post-hoc \ac{XAI} techniques on predictive models (e.g., classifying between controls and patients) to compute feature attribution scores (also referred to as saliency scores). Most recently, \ac{FM} have been proposed for studying brain dynamics \cite{zhou2025brain}. While previous models are designed for specific tasks, brain FMs (bFM) are capable of generalising to brain image analysis tasks, with or even without finetuning \cite{dong2024brain}.
However, their suitability for biomarker discovery remains uncertain as no existing studies have analysed the robustness of the salient features highlighted by these models. Existing bFM are often pre-trained on large open-source datasets dominated by healthy subjects 
and it remains to be seen whether this has a detrimental impact on the model's sensitivity to disorder states, which are often under-represented in such models.

For \ac{DL} models to be useful for biomarker discovery, they should identify a robust set of features that are reproducible across datasets. 
Despite high classification performance reported in studies using customised \ac{DL} models, existing research has revealed little convergence of salient features \cite{winter2022significance,abi2023candidate}. This suggests that these models are likely to have overfitted to noise and that common performance metrics such as accuracy, sensitivity, and specificity 
can only act as a basic but insufficient assessment of model performance. Thus, metrics beyond conventional measures, such as accuracies on test sets, are needed. One promising approach is to examine saliency scores generated by \ac{XAI} techniques.

The lack of robustness and large variations in existing saliency scores are attributed to three key factors: biological variations, data-based reasons, and model-dependent issues \cite{chan2025discovering}. 
Specifically, neurodevelopmental and neurodegenerative disorders are complex and often have heterogeneous presentations in both symptoms and nosology (e.g., subtypes) \cite{wu2023heterogeneity,kas2025towards}. Beyond biology-driven variations, it has been demonstrated that the choice of preprocessing pipelines impacts downstream performance \cite{luppi2024systematic}. 
Furthermore, even when using the same dataset and preprocessing pipeline, the choice of predictors and XAI algorithms (also known as attributors) also has a strong influence on feature salience scores, i.e., using different XAI algorithms could lead to different salient features being reported \cite{li2023classification,gallo2023functional}. 
Being exposed to much larger datasets than typical \ac{DL} models, \ac{FM} could potentially be more robust to such sources of variations, but techniques are needed to formally evaluate their robustness. 

Several evaluation metrics have been proposed in recent years. Early studies proposed to study the coherence of XAI algorithms by running various XAI algorithms on the same predictor \cite{li2023classification,gallo2023functional}, revealing that top features identified could vary even though the same predictor was used. In the case of brain dynamics, the temporal density of saliency maps has also been proposed as a means to observe unique temporal characteristics of each disorder's dynamic FC (dFC) \cite{rahman2022interpreting}. 

Expanding on our preliminary work \cite{girish2024robustness}, where we evaluated the robustness of salient FC features as potential biomarkers of brain disorders using static FC (sFC) metrics, we study the use of \ac{FM} for extracting dFC biomarkers of brain disorders - \ac{ASD}, \ac{ADHD} and \ac{AD} - from \ac{fMRI} datasets. Two novel dFC metrics are proposed to capture traits specific to brain dynamics that are not considered by existing evaluation metrics. Adding these metrics into the RE-CONFIRM framework, we evaluate existing bFM with publicly available code (BrainLM \cite{carobrainlm} and Brain-JEPA \cite{dong2024brain}), and compare them to existing state-of-the-art \ac{DL} models for dFC. 
We found that although bFM outperformed most existing \ac{DL} models, RE-CONFIRM metrics revealed that they capture hubs poorly as compared to specialised \ac{DL} models. Based on these insights, we propose Hub-LoRA, a novel finetuning technique for bFMs, that leverages brain hub structure to guide model parameter updates during training. \textcolor{black}{Across five datasets, we demonstrate that Hub-LoRA maintains model performance while selecting salient features that show better out-of-distribution generalization and greater overlap with regions reported in meta-analyses.}

Overall, our key contributions are: 
\begin{enumerate}
    \item We propose novel metrics for assessing the robustness of potential dFC biomarkers. Using these metrics, we found that existing bFM are more robust than non-FM in terms of data, labels, and parameters, but are less sensitive to functional hubs.
    \item We propose Hub-LoRA, a technique for finetuning bFM to improve their sensitivity to hubs. 
    \item \textcolor{black}{We demonstrated across five datasets, including two external validation datasets, that Hub-LoRA improves hub sensitivity in FMs and supports the identification of robust biomarkers.}
\end{enumerate}

\section{Methods}
In this section, we present RE-CONFIRM as a framework for evaluating the robustness of biomarkers detected by FM (see Figure \ref{fig:reconfirm}) and Hub-LoRA as an algorithm to finetune such models to enhance robustness. 

\subsection{RE-CONFIRM}
RE-CONFIRM identifies the best predictor-attributor combinations through a two stage pipeline. In the first stage, sanity checks derived from the Co-12 framework \cite{nauta2023anecdotal} filter out spurious combinations by assessing the sensitivity of feature attribution scores to variations in input data, labels, and model parameters. In the second stage, biomarker-specific metrics targeting top-$k$ salient features rank the surviving combinations. These include generic measures (e.g., Fidelity+) and connectome-aware measures (e.g., Hub Assortativity Coefficient).
The top-ranked combination is expected to yield better generalization as it captures the most relevant biomarkers. \textcolor{black}{In our preliminary work \cite{girish2024robustness}, we applied RE-CONFIRM to only GNN models trained on static FC, using metrics adapted from the Co-12 framework alongside two connectome-aware metrics: Modular ratio and HAC. These metrics do not account for the spatiotemporal structure of dFC. Moreover, removing salient features outright, as in Fidelity+, can result in inputs with inconsistent dimensionality. Here, we extend RE-CONFIRM to dFC and brain FMs by introducing two novel dFC-specific metrics: Hub Stability Index, which measures the temporal stability of brain hubs, and a spatio-temporal extension of Fidelity+ (SF+), which evaluates attribution faithfulness across spatial and temporal dimensions. These metrics evaluate the predictor-attributor combinations rather than the biological relevance of the biomarkers, which requires separate validation against meta-analyses and clinical measures.} 

\subsubsection{Stage I metrics} \mbox{}

Stage I metrics in RE-CONFIRM quantify differences between two sets of values: before and after a controlled change to the data or model. 
Changes in explanations are often quantified using the Hellinger distance. When $P$ and $Q$ are two discrete probability distributions obtained by discretising the distribution of saliency scores into a histogram of $B$ bins, the Hellinger distance is computed by the following equation:

\begin{equation}
    H(P, Q) = \frac{1}{\sqrt{2}} \sqrt{\sum_{i=1}^{B} \left( \sqrt{p_i} - \sqrt{q_i} \right)^2}.    
\end{equation}

\noindent where $p_i$ and $q_i$ are the empirical probability estimates of the $i$-th bin, obtained from samples $p \sim P$ and $q \sim Q$, respectively.

\textbf{Input Stability (IS):}
To evaluate the stability of an explanation, the model's ability to produce consistent outputs before and after input perturbations (e.g., adding Gaussian noise) is assessed. 
To compute this, we use a variant of Relative Input Stability \cite{agarwal2022rethinking}, which measures the relative distance between explanations of the original and perturbed input instances with respect to the distance between the input instances themselves. Let ${s} \in \mathbb{R}^{V\times V\times T}$ represent the vectorised representation of saliency scores across $V$ ROIs and $T$ time points for data sample ${x}$ and ${s}'$ represent the saliency scores generated from the perturbed data sample ${x}'$. IS is computed as:

\begin{equation}
    \text{IS}({x}, {x}', {s}, {s}') = \frac{\|{s} - {s}'\|_2}{\|{x} - {x}'\|_2}.
\end{equation}

\noindent where $\|\cdot\|_2$ denotes Euclidean norm. Higher values indicate greater instability, so lower values are preferred (Range [0, $+\infty$)).

\textbf{Label Randomization Check (LRC):}
LRC quantifies the sensitivity of explanations to the relationship between input data and labels. The labels in the training dataset are randomly permuted, thereby breaking the true input–label relationship, and a new model is trained on this randomized dataset. Explanations are then generated for the same inputs using both the model trained on true labels (\(P_{\theta^*}\)) and the model trained on randomized labels (\(Q_{\tilde{\theta}}\)). If the explanations are sensitive to the input–label relationship, they should differ significantly between the two models, leading to a larger Hellinger distance. Higher LRC values are desired (close to 1, Range [0,1]). LRC is represented as:

\begin{equation}
LRC = H(P_{\theta^*}, Q_{\tilde{\theta}})
\end{equation}

\textbf{Model Parameter Randomization Check (MPRC):}
This method evaluates the faithfulness and sensitivity of explanations to the model's learned parameters. Model weights perturbed via reinitialization, destroying the learned representations. The explanations from the perturbed model (\(Q'_{\tilde{\theta}}\)) are then compared with those of the original trained model (\(P_{\theta^*}\)). If the explanations remain unchanged (i.e. low Hellinger distance) despite these perturbations, it indicates that they do not reflect the underlying reasoning of the model. Higher MPRC values are desired (close to 1, Range [0,1]). MPRC is given by:

\begin{equation}
MPRC = H(P_{\theta^*}, Q'_{\tilde{\theta}})
\end{equation}

\noindent \textcolor{black}{Stage I is the filtering stage, where we retain a predictor–attributor combination only if it passes all three metric sanity checks: LRC, MPRC, and IS. Combinations are excluded if their explanations are insensitive to label or parameter randomization (LRC or MPRC close to zero) or highly unstable (IS markedly higher than that of other combinations).}

\subsubsection{Stage II metrics} \mbox{}

\textbf{Hub Stability Index (HSI):}
\textcolor{black}{Existing studies show that the most dynamic functional connections are concentrated at brain hubs \cite{zalesky2014time}. In contrast, whole-brain connectivity states persist over extended periods, and the time spent in these states is altered in ASD and ADHD \cite{watanabe2017brain, shappell2021children}.} Motivated by these observations, we propose HSI to quantify the temporal consistency of the proportion of network hubs. The key idea is that the model's ability to capture hub dynamics, as measured through its saliency scores $S \in \mathbb{R}^{N \times V \times V \times T}$, where \(N\) denotes the number of subjects, should at least match the variance of the proportion of hubs as computed from the dFC data $X \in \mathbb{R}^{N \times V \times V \times T}$.
Since saliency scores can be noisy at its lowest granularity (e.g. pixels \cite{kim2019saliency} and in this context, time points), a less direct approach is designed in contrast to HAC.

At each sliding time window, we compute the fraction of nodes identified as hubs that are defined by ranking nodes according to their ambivert degree \cite{gupta2020ambivert}. This produces a time series of hub proportions. Temporal variability in hub expression is quantified by computing the variance of the sequence. To allow comparison across scans with different numbers of time points, the variance is subsequently normalized, which reduces the influence of extreme values at individual time points, yielding a bounded variability measure. HSI is then defined as one minus this normalized variance, such that higher HSI values (close to 1) indicate greater consistency of hub expression over time, whereas lower values (close to 0) indicate instability. HSIs are computed separately for hubs derived from \ac{FC} values and from saliency scores. We then combine the two using their harmonic mean to obtain the overall combined stability score. \textcolor{black}{This metric assesses whether the temporal stability of hub proportions derived from saliency scores aligns with that derived from \ac{FC} values, focusing on overall hub prevalence rather than the identities of individual hubs.} Algorithm 1 provides a summary of how HSI is computed.

\begin{algorithm}[!ht]
\begin{algorithmic}[1]
\caption{Hub Stability Index for \(X\)}

\State \textbf{Input:} dFC values \(X\)

\State \textbf{Output:} $\text{Hub Stability Index, } HSI$ 
\State Identify hub nodes \(\mathcal{H}_t\) and compute proportion $P_{\mathcal{H}}(t)$ from each sliding window

\For{$t = 1,2,3,...T$}
    \State $\mathcal{H}_t = \emptyset$
    \For{$v = 1,2,3,...V$}
        \State $\mathcal{H}_t \leftarrow \mathcal{H}_t \cup \{ v \in V, Ambivert(v) > 1\}$
    \EndFor
    \State $P_{\mathcal{H}}(t) = \frac{|\mathcal{H}_t|}{V}$ \Comment{Proportion of nodes being hubs}
\EndFor

\State Compute variance and perform normalization to facilitate downstream comparisons
\State $\sigma^2_{\mathrm{norm}} = \frac{\mathrm{Var}(P_{\mathcal{H}}(t))}{\max(P_{\mathcal{H}}(t))^2 + \epsilon}$ \Comment{$\epsilon$ to prevent division by 0} 

\State $HSI = 1 -  \sigma^2_{\mathrm{norm}}$
    
\end{algorithmic}
\end{algorithm}

After computing hub stability for both $X$ and $S$, we compute the harmonic mean of $\mathrm{HSI}_X$ and $\mathrm{HSI}_S$ to arrive at the overall HSI value:
\textcolor{black}{
\begin{equation}
     \mathrm{HSI_{overall}} = \frac{2 \times \mathrm{HSI}_X \times \mathrm{HSI}_S}{\mathrm{HSI}_{X} + \mathrm{HSI}_S}.
\end{equation}
}

\textbf{Spatio-temporal Fidelity+ (SF+):}

Fidelity quantifies the faithfulness of explanations to a model's predictions by measuring changes in predicted probabilities after selective feature ablation. Specifically, it compares the original predictions with those obtained after removing either the top \emph{k} most important features (Fidelity+) or the bottom \emph{k} least important features (Fidelity$-$). Top $k$ features are often reported in DL-driven biomarker discovery, thus Fidelity+ (F+) is an intuitive method to assess the importance of these top features. 

The underlying premise of Fidelity+ \cite{yuan2022explainability} is that if the selected subset of features are indeed important, its removal should lead to a significant change in the model's predictions. However, this approach is not well suited to spatio-temporal settings, as strictly removing top \emph{k} spatio-temporal features not only result in inputs with inconsistent dimensionalities across time and space (making model training infeasible without zero-padding, which has been argued to introduce an out-of-sample setting and thus be inappropriate \cite{zheng2024f}), but also has little biological meaning especially when the salient features are widely spread across features and time points. 

To address this, we propose SF+ as an alternative metric with better biological grounding. In SF+, a sliding window of length $t'$ is applied along the temporal dimension to aggregate saliency scores across time. Based on the aggregated scores, the top \emph{k} features are pruned. The resulting missing values are then imputed in the ablated spatio-temporal feature matrix using the K-Nearest Neighbor (K-NN) method (K not to be confused with the top $k$ features). Both $k$ and $t'$ are treated as tunable hyperparameters. This can be expressed via the following equation:
\begin{equation}
    SF+ = \frac{1}{N} \sum_j (f(X_j) - f(X_j^+)).
\end{equation}
\textcolor{black}{where $f(\cdot)$ denotes the model's predicted probability for its originally predicted class}, $X_j$ represents the spatio-temporal \ac{FC} matrix of subject $j$ from a dataset of $N$ subjects, and $X_j^+$ represents the spatio-temporal FC matrix of subject $j$ with the top $k$ saliency node features removed across time based on our proposed sliding window approach. Higher SF+ value is desired (close to 1.0, \textcolor{black}{Range [-1,1]}).

\noindent \textcolor{black}{Stage II is the ranking stage, where surviving combinations are ranked using static FC-based metrics (Fidelity+ and HAC) and dynamic FC-based metrics (SF+ and HSI). The final ranking is determined by the mean rank across these metrics, with accuracy used to break ties.}

\subsection{Hub-LoRA algorithm}

Brain functional connectivity hubs are often implicated in many neurological disorders, such as \ac{ASD} and \ac{ADHD} \cite{power2013evidence, wang2019functional}. Since brain FMs are largely pre-trained on healthy controls, we hypothesise that vanilla finetuning in the form of linear probing or full finetuning is insufficient to guarantee that hubs are captured. This is especially so for small and high-dimensional datasets like dFC as models trained on them are prone to underspecification \cite{ji2021convolutional}. 

To address this issue, we propose Hub-LoRA as a finetuning method for DL models grounded in LoRA reparametrization \cite{hu2022lora}. LoRA introduces low-rank trainable matrices to update pre-trained weights while keeping the original weights frozen. Hub-LoRA extends this formulation by applying the adaptation to the first layer of the neural network, allowing the low-rank update to directly interact with the input data and influence the feature selection process. 

Let $W^\text{new}$ represent the new weights after fine-tuning the original pre-trained weight matrix $W$. 
\begin{equation}
W^\text{new} = W + \Delta W \quad\quad\quad \Delta W = B A
\end{equation}
where weight update \(\Delta W\) is decomposed into two low-rank matrices $A \in \mathbb{R}^{r \times k'}$, $B \in \mathbb{R}^{k'' \times r}$, $r$ represents the chosen rank of the matrices, $k'$ represents the number of input features (typically ROIs) and $k''$ represents the size of the output. During finetuning, $W$ is frozen while $A$ and $B$ are trained in their decomposed form. 
When applied to the first layer, which takes in the input matrix $X \in \mathbb{R}^{N\times k'}$, $A$ plays the role of choosing which sets of features to adjust weights for. 

Instead of initialising it randomly and freezing it like in LoRA-FA \cite{zhang2023lora}, we measure the hubness of each node using the ambivert degree and follow a threshold of 1.0 following recommendations by existing literature \cite{sigar2023altered} to ascertain whether a node is considered a hub. Then, for every hub node, a one-hot vector corresponding to the index of the node is created as a row in $A$, i.e., rank $r$ is set as the number of hubs. These values are frozen throughout the finetuning process, allowing the model to remain focused on these hub nodes only when adjusting the weight matrix $W$.

$B$ plays the role of feature scaling, which is implemented as a vector of scaling factors ${d} \in \mathbb{R}^r$, which is defined as the ranking differences between the existing saliency score vector (derived from the previously finetuned model) and the vector of ambivert degrees. Ranking here refers to the ordered position relative to other values in the vector, as ordered based on its value. It is not to be confused with the matrix rank $r$, but $r$ is indeed the length of the ranking vector defined below.

Factors ${d}$ are computed by the following steps. Let ${u} \in \mathbb{R}^r$ represent the attribution scores and ${a} \in \mathbb{R}^r$ represent the ambivert scores for the $r$ hub nodes. First, each vector is converted into its corresponding ranking vector ${r}_u$ and ${r}_a$. For each hub node $v \in \{1, \dots, r\}$, the ranking is defined as:
\begin{equation}
    [\pi({u})]_v = \# \{ j: {u}_j \leq {u}_v \}, \quad [\pi({a})]_v = \# \{ j: {a}_j \leq {a}_v \}.
\end{equation}
Then, the rank differences are computed via the squared rank distance:
\begin{equation}
    {d}_v = ([\pi({u})]_v - [\pi({{a}})]_v)^2.
\end{equation}
This gives a good initialization for the model to determine which hub nodes to focus on. ${d}$ is fixed throughout finetuning and each column in $B'$ is scaled element-wise by the corresponding entry in ${d}$. Thus, $B'$ 
is the only learnable component in Hub-LoRA:

\begin{equation}
B = B' \odot {d}.
\end{equation}

Overall, Hub-LoRA narrows the model's focus on hub nodes and provides targeted updates specifically to these nodes with discordant rankings. Algorithm 2 provides a summary of the steps taken to compute the updated weights $W^\text{new}$.

\begin{algorithm}[!ht]
\begin{algorithmic}[1]
\caption{Hub-LoRA}

\State \textbf{Input:} Pre-trained weight matrix \(W \in \mathbb{R}^{k'' \times k'}\), low-rank matrices \(A, B'\), saliency score vector \({u} \in \mathbb{R}^{r}\), ambivert degree vector \({a} \in \mathbb{R}^{r}\)

\State \textbf{Output:} Updated weight matrix \(W'\)
\State Freeze W and \(A \in \mathbb{R}^{r \times k'}\)
\State Compute rank vectors and squared rank distance
\For{$v = 1,2,3,...r$}
    \State $[\pi({u})]_v = \# \{ j: {u}_j \leq {u}_v \}$
    \State $[\pi({a})]_v = \# \{ j: {a}_j \leq {a}_v \}$
    \State ${d}_v = ([\pi({u})]_v - [\pi({{a}})]_v)^2$
\EndFor

\State Set \(B = B' \odot {d}\), where \(B' \in \mathbb{R}^{k'' \times r}\) is trainable

\State \textbf{Return} $W' = W + (B' \odot {d}) \cdot A$
\end{algorithmic}
\end{algorithm}

\section{Experiments and Results}

\subsection{Datasets}

The ABIDE I dataset \cite{di2014autism} contains 387 resting-state \ac{fMRI} scans from individuals diagnosed with \ac{ASD} and 436 typically developing controls, collected from 20 sites. We also used the ABIDE II dataset \cite{di2017enhancing}, which comprises 127 individuals with ASD and 131 controls from 4 sites (BNI, EMC, GU, and IP). Data from the ADHD-200 dataset \cite{brown2012adhd} and CNI-TLC 
\cite{schirmer2021neuropsychiatric} were used to further validate our findings. ADHD-200 contains rs-fMRI scans from 301 subjects diagnosed with \ac{ADHD} and 549 age-matched controls, collected from 6 sites (NI, NYU, OHSU, PKU, KKI, and WUSTL). Preprocessed resting-state \ac{fMRI} data were downloaded from the Preprocessed Connectome Project. Data from the C-PAC pipeline and Athena pipeline were selected for ABIDE (both I and II) and ADHD-200, respectively. CNI-TLC comprised 100 individuals diagnosed with ADHD and 100 controls. For AD, we ran experiments on the ADNI dataset \cite{petersen2010alzheimer}, using 100 subjects diagnosed with AD and 100 healthy controls.

Following recommended practices \cite{leonardi2015spurious}, we constructed dFC matrices by segmenting the BOLD signals into overlapping windows, each 60s in length with a 1s stride. Power atlas \cite{power2011functional} was used to identify 264 diverse regions of interest. For biomarker analysis, as anatomical labels for ROIs are not provided in the Power atlas, we adopted the anatomical labels from the Crossley atlas \cite{crossley2013cognitive}. Each Power ROI is mapped to a Crossley ROI by assigning the nearest anatomical label based on Euclidean distance between ROI coordinates. We computed the mean time series of all voxels within a sphere of radius 2.5 mm around each ROI. FC matrices were computed by determining the Pearson correlation between the mean activation time series for each ROI pair. The dFC matrices generated from each sliding window were subsequently used as model inputs.

\subsection{Implementation Details}

To assess the robustness of dFC biomarkers derived from FMs, we performed experiments using five different seeds on several predictors, comparing FMs (BrainLM \cite{carobrainlm}, Brain-JEPA \cite{dong2024brain}) with non-FMs (STGCN \cite{gadgil2020spatio}, STAGIN \cite{kim2021learning}, MSC-Mamba \cite{li2025multiscale}).
To enable comparisons across different brain atlases, all atlases were first transformed into a common reference space to ensure spatial alignment. The AAL-424 atlas 
(BrainLM), AAL-116 
(MSCMamba), and Schaefer-400 atlas 
(BrainJEPA, STAGIN) were converted into the Power atlas by using coordinate-based mapping in MNI space. For each region in the source atlas, the Euclidean distance (the straight-line distance in 3D space) to every brain node was computed, and the node with the smallest distance was assigned as its closest match. This ensures that each anatomical or functional region is consistently mapped to a sphere, enabling effective cross-model comparisons. 

For the FMs, we explored various finetuning approaches to enable their applications for downstream tasks. Since model weights of FMs are provided based on pre-training and not for downstream tasks, we first attached a classification head and trained only the classifier by minimising the cross-entropy loss function while keeping the pre-trained model weights frozen. This is also known as linear probing. For subsequent finetuning, we experimented with a simple approach of tuning all model parameters in the \ac{FM} and more specific approaches based on LoRA, performed on top of linear probing. 
For full finetuning, we followed the recommendations and default hyperparameters provided in the respective papers. Specifically, we increased number of MLP heads from 1 to 3 for BrainLM. BatchNorm was added to account for normalization. For Brain-JEPA, AdamW was used as the optimiser and the number of epochs and batch size was reduced from 50 to 30 and from 16 to 4, respectively. For LoRA finetuning, it was applied on the attention layers in the encoder for BrainLM and Brain-JEPA and on the linear layer of MSC-Mamba.

Saliency scores for identifying potential biomarkers were obtained via two approaches. First, attention scores from the attention layers of each architecture were retrieved across all seeds. Scores were computed for each sliding time-window segment across all models, then averaged across seeds. 
Second, for post-hoc explainability, Integrated Gradients \cite{sundararajan2017axiomatic} was used to compute saliency scores from each model. The baseline was defined as the mean input computed across all controls in the training dataset and default parameters were used (e.g. \texttt{n\_steps} was set as 50). 

On top of accuracy measures, we utilised the five metrics presented in RE-CONFIRM. Models are first compared based on predictive accuracy. Among the better-performing models, Stage I metrics (IS, LRC, MPRC) are then used for filtering. Finally, Stage II metrics (SF+ and HSI), which are neuroscience-informed, are applied for further evaluation.  The number of neighbours used for nearest neighbour imputation in SF+ was set to 5, the stride was set to 1s and $k$ was set as 20, but was also varied in sensitivity analyses shown below. In the following tables, we further include baseline metrics such as Fidelity+ \cite{nauta2023anecdotal} and Hub Assortativity Coefficient (HAC) \cite{girish2024robustness} to contrast them with SF+ and HSI, respectively.

\subsection{Results}

\subsubsection{Classification accuracies}

\begin{table}[!ht]
\caption{Comparison of classification accuracies (\%) and number of trainable model parameters. \label{tab:accuracies}}
\centering
\resizebox{\columnwidth}{!}{
\begin{tabular}{llllll}
\toprule
Models & Type & ABIDE & ADHD & ADNI & Params \\
\midrule
BrainGNN & sFC & 61.34 \(\pm\) 4.57 & 57.71 \(\pm\) 5.41 & 58.23 \(\pm\) 4.94 & 93k \\
\midrule
STGCN & \multirow{4}{*}{dFC} & 59.82 \(\pm\) 5.69 & 58.24 \(\pm\) 6.23 & 60.82 \(\pm\) 5.10 & 208k \\
STAGIN &  & 71.97 \(\pm\) 3.44 & 66.88 \(\pm\) 2.80 & 72.44 \(\pm\) 4.78 & 1,004k \\
MSC-Mamba &  & 72.44 \(\pm\) 3.80 & 70.58 \(\pm\) 3.49 & 74.19 \(\pm\) 3.99 & 1,605k \\
BRAINMAP &  & 73.18 \(\pm\) 2.32 & \textbf{74.04 \(\pm\) 2.10} & \textbf{76.08 \(\pm\) 3.51} & 1,026k \\
\midrule
BrainLM & \multirow{2}{*}{bFM} & 73.22 \(\pm\) 3.46 & 71.77 \(\pm\) 3.30 & 74.86 \(\pm\) 3.37 & 13,020k \\
Brain-JEPA &  & \textbf{74.05 \(\pm\) 4.14} & 72.08 \(\pm\) 2.01 & 75.80 \(\pm\) 3.28 & 86,900k  \\
\bottomrule
\end{tabular}}
\end{table}

Table \ref{tab:accuracies} presents results where non-FM models (including both sFC and dFC models) were trained on the datasets and linear probing was performed on FMs (i.e. pre-trained weights are frozen and only the classifier head was tuned). Brain-JEPA has the highest performance on the ABIDE dataset, but this is largely statistically insignificant when compared against other \ac{FM} or the best non-FM (BRAINMAP), based on Welch's t-test performed across five independent runs (p-value = 0.69 for BRAINMAP, p-value = 0.74 for BrainLM). On the ADHD and ADNI dataset, BRAINMAP was the best-performing model, achieving significant improvements over most models, except the bFMs.

One notable variation in these experiments is the trainable number of parameters involved, which is often not well controlled for in existing studies. However, we found that our findings were still valid even after increasing the number of trainable parameters of BrainGNN and STGCN to 1 million parameters (Table \ref{tab:accuracies-additional}). Since these two models consistently underperformed, they were excluded from further analyses.

\begin{table}[ht]
\caption{Classification accuracies after finetuning with LoRA
\label{tab:acc-finetune}}
\centering
\resizebox{\columnwidth}{!}{
\begin{tabular}{l|lll|ll}
\toprule
Layers & MSC-Mamba & BrainLM & Brain-JEPA & STAGIN & BRAINMAP \\
\midrule
ABIDE\\
\midrule

One & 72.80\(\pm\)3.37 & 74.12\(\pm\)3.04 & 75.20\(\pm\)3.82 & 72.29\(\pm\)2.49 & 73.93\(\pm\)1.86 \\

Half & 73.38\(\pm\)2.86 & 74.86\(\pm\)3.22 & 76.10\(\pm\)3.15 & \textbf{72.91\(\pm\)2.93} & 74.32\(\pm\)2.11 \\

All & \textbf{74.40\(\pm\)3.39} & \textbf{75.35\(\pm\)3.09} & \textbf{77.10\(\pm\)2.88} & 72.78\(\pm\)2.52 & \textbf{74.84\(\pm\)1.91} \\

Full & 72.44\(\pm\)3.80 & 73.22\(\pm\)3.46 & 74.05\(\pm\)4.14 & 71.97\(\pm\)3.44 & 73.18\(\pm\)2.32 \\

\midrule
ADHD-200\\
\midrule

One & 72.02\(\pm\)3.35 & 73.04\(\pm\)3.14 & 73.42\(\pm\)2.16 & 67.47\(\pm\)2.94 & 74.70\(\pm\)1.92 \\

Half & 72.27\(\pm\)2.92 & 73.88\(\pm\)3.12 & 75.12\(\pm\)3.07 & \textbf{67.71\(\pm\)2.73} & 75.01\(\pm\)2.16 \\

All &  \textbf{73.60\(\pm\)3.51} &  \textbf{75.54\(\pm\)3.18} &  \textbf{76.34\(\pm\)3.04} & 67.53\(\pm\)2.25 & \textbf{75.44\(\pm\)2.22} \\

Full & 70.58\(\pm\)3.49 & 71.77\(\pm\)3.30 & 72.08\(\pm\)2.01 & 66.88\(\pm\)2.80 & 74.04\(\pm\)2.10 \\

\bottomrule
\end{tabular}}
\end{table}

These baseline results can be further improved through finetuning using parameter-efficient techniques. In this study, we adopt the original LoRA formulation and do not adapt bias terms, leaving them frozen during finetuning. LoRA is applied to fully connected layers and convolutional layers. For attention layers, the query, key, and value projections are treated as a single combined matrix, while the output projection remains frozen. We found that full finetuning only results in slight improvements that were not significant. As seen in Table \ref{tab:acc-finetune} \footnote[1]{\textcolor{black}{"One" indicates LoRA is applied to the first layer of the model, "Half" to the first half of the layers, and "All" to every model layers. "Full" indicates that all model parameters are fine-tuned without LoRA.}}, when finetuning was done in a more controlled fashion via LoRA, simply finetuning one layer (first layer) results in a model performance on par with the fully tuned model. This is consistently observed throughout both \ac{FM} and non-FM settings and across all datasets (ADNI not shown due to space constraints).

\begin{table}[ht]
\caption{Classification accuracies after varying LoRA ranks
\label{tab:acc-ranks}}
\centering
\resizebox{\columnwidth}{!}{
\begin{tabular}{l|lll|ll}
\toprule
Rank & MSC-Mamba & BrainLM & Brain-JEPA & STAGIN & BRAINMAP \\
\midrule
ABIDE\\
\midrule
1 & 74.40\(\pm\)3.39 & 75.35\(\pm\)3.09 & 77.10\(\pm\)2.88 & 72.91\(\pm\)2.93 & 74.84\(\pm\)1.91\\
2 & 74.68\(\pm\)2.78 & 75.50\(\pm\)2.60 & 77.67\(\pm\)3.03 & \textbf{73.28\(\pm\)2.44} & 75.03\(\pm\)2.94\\
4 & \textbf{74.77\(\pm\)2.88} & \textbf{75.67\(\pm\)2.31} & 77.92\(\pm\)2.74 & 72.80\(\pm\)2.12 & \textbf{75.18\(\pm\)2.66}\\
8 & 74.68\(\pm\)2.78 & 75.56\(\pm\)2.73 & \textbf{78.16\(\pm\)2.82} & \textbf{73.28\(\pm\)2.44} & 74.91\(\pm\)2.03\\

\midrule
ADHD-200\\
\midrule

1 & 73.60\(\pm\)3.51 & 75.54\(\pm\)3.18 & 76.34\(\pm\)3.04 & 67.71\(\pm\)2.73 & 75.44\(\pm\)2.22\\
2 & \textbf{74.11\(\pm\)2.63} & 75.83\(\pm\)2.87 & 77.12\(\pm\)2.85 & \textbf{67.84\(\pm\)2.58} & 75.23\(\pm\)2.19\\
4 & 73.89\(\pm\)2.77 & \textbf{76.08\(\pm\)2.46} & \textbf{78.04\(\pm\)2.96} & 67.10\(\pm\)2.21 & 75.53\(\pm\)3.03\\
8 & 73.42\(\pm\)2.64 & 75.72\(\pm\)2.15 & 77.84\(\pm\)2.72 & 67.26\(\pm\)2.33 & \textbf{75.72\(\pm\)3.08}\\
\bottomrule
\end{tabular}}
\end{table}

Finetuning via LoRA involves several design considerations, such as the rank used and the layer types on which LoRA is applied. Table \ref{tab:acc-ranks} reveals that using a rank of 2 or 4 suffices in most cases. As for layer types, we did not find any significant differences in model performances when the choice of layers was varied (Conv, linear for MSC-Mamba; Attention, Linear for STAGIN; linear, conv, attention for BrainMAP; attention for BrainLM and Brain-JEPA). Since STAGIN and MSC-Mamba had the lowest performance across all models, they were excluded from subsequent analyses, and for those that are included, the models with the best rank were used.

\subsubsection{RE-CONFIRM metrics}

\begin{table}[ht]
\caption{RE-CONFIRM evaluation metrics for IG attributor.
\label{tab:metrics-ig}}
\centering
\resizebox{\columnwidth}{!}{
\begin{tabular}{l|lll|lll}
\toprule
 & \multicolumn{3}{c|}{ABIDE} & \multicolumn{3}{c}{ADHD-200} \\
\cmidrule(lr){2-4} \cmidrule(lr){5-7}
Metrics & BrainLM & Brain-JEPA & BRAINMAP & BrainLM & Brain-JEPA & BRAINMAP \\
\midrule
IS ($\downarrow$) & \textbf{2.86\(\pm\)1.35} & 3.12\(\pm\)1.74 & 4.05\(\pm\)1.23 & \textbf{2.98\(\pm\)1.37} & 3.21\(\pm\)1.91 & 4.22\(\pm\)1.72 \\
LRC ($\uparrow$) & 0.48\(\pm\)0.02 & \textbf{0.54\(\pm\)0.02} & 0.51\(\pm\)0.03 & 0.56\(\pm\)0.02 & \textbf{0.57\(\pm\)0.03} & 0.52\(\pm\)0.02 \\
MPRC ($\uparrow$) & 0.53\(\pm\)0.02 & \textbf{0.58\(\pm\)0.02} & 0.55\(\pm\)0.04 & 0.57\(\pm\)0.01 & \textbf{0.62\(\pm\)0.02} & 0.58\(\pm\)0.02\\
\midrule
SF+ ($\uparrow$) & 0.52\(\pm\)0.02 & \textbf{0.56\(\pm\)0.01} & 0.50\(\pm\)0.02 & 0.53\(\pm\)0.03 & \textbf{0.57\(\pm\)0.02} & 0.52\(\pm\)0.02 \\
HSI ($\uparrow$) & 0.67\(\pm\)0.03 & 0.70\(\pm\)0.03 & \textbf{0.77\(\pm\)0.03} & 0.68\(\pm\)0.02 & 0.72\(\pm\)0.03 & \textbf{0.76\(\pm\)0.02}\\
\midrule
F+ ($\uparrow$) & 0.53\(\pm\)0.02 & \textbf{0.55\(\pm\)0.03} & 0.52\(\pm\)0.02 & 0.52\(\pm\)0.01 & \textbf{0.54\(\pm\)0.02} & 0.44\(\pm\)0.02\\
HAC ($\uparrow$) & 0.29\(\pm\)0.01 & 0.28\(\pm\)0.01 & \textbf{0.34\(\pm\)0.02} & 0.26\(\pm\)0.02 & 0.26\(\pm\)0.01 & \textbf{0.36\(\pm\)0.03} \\
\bottomrule
\end{tabular}
}
\end{table}

Looking at the RE-CONFIRM metrics, it is evident from Table \ref{tab:metrics-attention} and Table \ref{tab:metrics-ig} that the FMs outperformed BRAINMAP (non-FM) in terms of IS, LRC and MPRC. This is seen across all datasets (ADNI not shown due to space limitations) and suggests that FMs have more robust model performances for generic metrics.
However, biomarker-specific metrics suggest that BRAINMAP is more sensitive to hubs than the FMs. Additionally, while BRAINMAP has lower Fidelity+, SF+ is similar to the FMs, suggesting that the removal of top-$k$ spatio-temporal features identified by each model results in a similar impact on model performance. 
These findings are also consistently observed when attention is used as the attributor (instead of IG). However, comparison of IS values revealed that IG scores have much better IS than attention.
Overall, our results reveal that FMs do indeed provide a boost in terms of model robustness, but the salient features identified by FMs, even after finetuning, do not seem to capture hubs as well as non-FM approaches that are designed for such purposes.

\subsubsection{Hub-LoRA}

\begin{table}[ht]
\caption{Evaluation metrics of Brain-JEPA (All) with LoRA and Hub-LoRA finetuning
\label{tab:metrics-hublora}}
\centering
\resizebox{\columnwidth}{!}{
\begin{tabular}{l|ll|ll|ll}
\toprule
 & \multicolumn{2}{c|}{ABIDE} & \multicolumn{2}{c|}{ADHD-200} & \multicolumn{2}{c}{ADNI} \\
\cmidrule(lr){2-3} \cmidrule(lr){4-5} \cmidrule(lr){6-7}
Metrics & LoRA & Hub-LoRA & LoRA & Hub-LoRA & LoRA & Hub-LoRA \\
\midrule
Acc ($\uparrow$) & 78.2\(\pm\)2.8 & \textbf{79.2\(\pm\)1.9} & 78.0\(\pm\)3.0 & \textbf{79.0\(\pm\)2.1} & 77.2\(\pm\)3.1 & \textbf{77.9\(\pm\)3.1} \\
\midrule
IS ($\downarrow$) & 3.12\(\pm\)1.74 & \textbf{2.75\(\pm\)1.74} & 3.21\(\pm\)1.91 & \textbf{2.85\(\pm\)1.79} & 3.76\(\pm\)1.74 & \textbf{3.35\(\pm\)1.58}  \\
LRC ($\uparrow$) & \textbf{0.54\(\pm\)0.02} & \textbf{0.54\(\pm\)0.02} & \textbf{0.57\(\pm\)0.03} & 0.56\(\pm\)0.02 & \textbf{0.53\(\pm\)0.03} & 0.51\(\pm\)0.02  \\
MPRC ($\uparrow$) & \textbf{0.58\(\pm\)0.02} & 0.57\(\pm\)0.02 & \textbf{0.62\(\pm\)0.02} & 0.58\(\pm\)0.02 & 0.56\(\pm\)0.02 & \textbf{0.57\(\pm\)0.01}\\
\midrule
SF+ ($\uparrow$) & 0.56\(\pm\)0.01 & \textbf{0.59\(\pm\)0.01} & \textbf{0.57\(\pm\)0.02} & \textbf{0.57\(\pm\)0.02} & 0.54\(\pm\)0.03 & \textbf{0.58\(\pm\)0.03} \\
HSI ($\uparrow$) & 0.70\(\pm\)0.03 & \textbf{0.74\(\pm\)0.02} & 0.72\(\pm\)0.03 & \textbf{0.73\(\pm\)0.01} & 0.71\(\pm\)0.01 & \textbf{0.73\(\pm\)0.02} \\
\midrule
F+ ($\uparrow$) & 0.55\(\pm\)0.03 & \textbf{0.56\(\pm\)0.03} & \textbf{0.54\(\pm\)0.02} & \textbf{0.54\(\pm\)0.02} & 0.52\(\pm\)0.02 & \textbf{0.55\(\pm\)0.03} \\
HAC ($\uparrow$) & 0.28\(\pm\)0.01 & \textbf{0.31\(\pm\)0.01} & 0.26\(\pm\)0.01 & \textbf{0.32\(\pm\)0.02} & 0.29\(\pm\)0.03 & \textbf{0.32\(\pm\)0.02} \\
\bottomrule
\end{tabular}
}
\end{table}

Table \ref{tab:metrics-hublora}\footnotemark[1] shows how the introduction of Hub-LoRA influenced both classification accuracies and RE-CONFIRM metrics. Although additional finetuning of the Brain-JEPA (all) model via Hub-LoRA resulted in improvements that are not statistically significant based on Welch's t-test 
\textcolor{black}{(p-value = 0.53 for ABIDE, 0.56 for ADHD, 0.73 for ADNI), it led to a modest increase in HAC for Brain-JEPA and reduced the gap from BRAINMAP in both HSI and HAC.} These improvements were achieved without any significant sacrifices in other evaluation metrics, and findings are consistently observed in other models (e.g. applying Hub-LoRA to Brain-JEPA (full), not shown due to space constraints).

\subsubsection{Biomarker analysis}

Biomarker analysis was performed under four settings: BRAINMAP (best non-FM), Brain-JEPA (Full), Brain-JEPA (All) and Hub-LoRA on Brain-JEPA (All). This allows us to compare between non-FM and \ac{FM} as well as to investigate the effects of various finetuning techniques on biomarker discovery, including Hub-LoRA. Due to space constraints, visualisations are limited to ABIDE and ADHD.

\begin{figure}[t]
   \centering
   \includegraphics[width=0.95\linewidth]{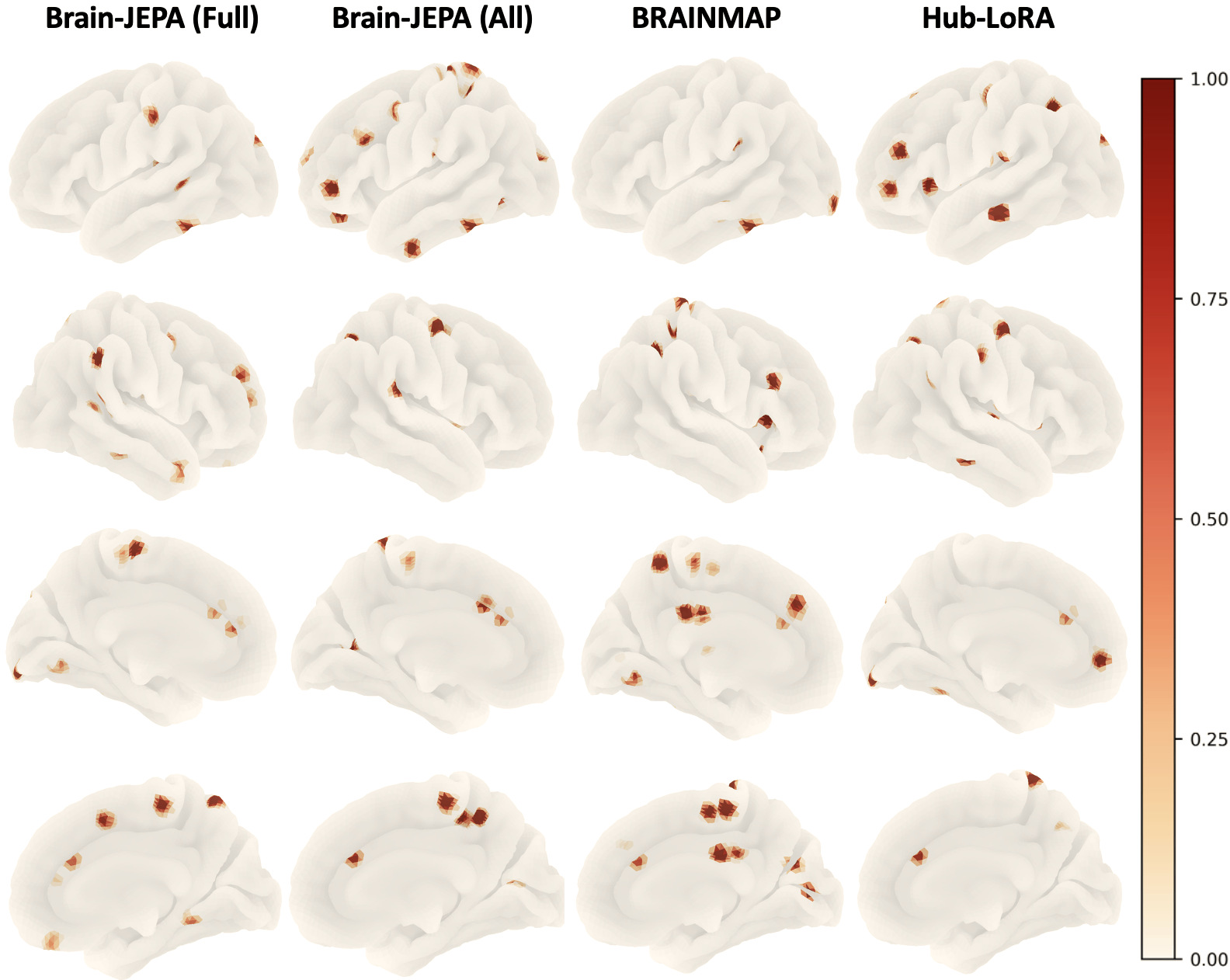}
   \caption{Surface plots based on normalized attributions produced via IG, for ABIDE. Rows: Left lateral, Right lateral, Right medial, Left medial.}
   \label{fig:ig_abide}
\end{figure}

When attention scores are used (Figure \ref{fig:attn}), potential ASD biomarkers highlighted by Brain-JEPA (full) include middle occipital gyrus, parietal lobule and regions such as claustrum, extra-nuclear and fusiform gyrus. 
While Brain-JEPA (all) also highlighted the middle occipital gyrus, it is more focused on both precentral and postcentral gyrus, as well as the frontal gyrus (medial, inferior, superior, and middle). 
BRAINMAP contains a mix of regions, including the parietal lobule, middle and superior frontal gyrus, and has an additional focus on ROIs from the salience network, such as the cingulate gyrus and insula, as well as the lingual gyrus. 

When a different attributor is used (IG, Figure \ref{fig:ig_abide}\footnotemark[1]), there is a remarkable change in the salient features identified by the models. 
Brain-JEPA (Full) no longer focuses on the occipital gyrus and parietal lobe, instead identifying ROIs related to cognition as important. This includes the medial and middle frontal gyrus, anterior cingulate, precuneus, declive and supramarginal gyrus.
While Brain-JEPA (All) still identified the superior frontal gyrus as an important ROI, it also has an eclectic mix of ROIs, including the lingual gyrus, insula, fusiform gyrus, precentral gyrus, paracentral lobule, and inferior parietal lobule.
BRAINMAP retains its focus on ROIs from the salience network and frontal gyrus but now extends to the fusiform gyrus and subcortical regions (thalamus). Notably, these ROIs are closer to what existing literature reveals \cite{gupta2020ambivert,guo2024systematic}. It also includes the paracentral lobule, inferior parietal lobe and postcentral gyrus. 
When Hub-LoRA is applied to Brain-JEPA (All), prominent hubs implicated in ASD, such as the fusiform gyrus and cingulate gyrus, are highlighted, providing qualitative support to the higher HAC observed in Table \ref{tab:metrics-hublora}. 
\textcolor{black}{Overall, BRAINMAP-IG combination largely recovers canonical networks implicated in ASD (default mode network (DMN), salience network, thalamocortical, fusiform) and Hub-LoRA helped Brain-JEPA (All) to recover more neurobiologically \textcolor{black}{plausible} biomarkers such as cingulate gyrus.} 

\begin{figure}[t]
   \centering
   \includegraphics[width=0.95\linewidth]{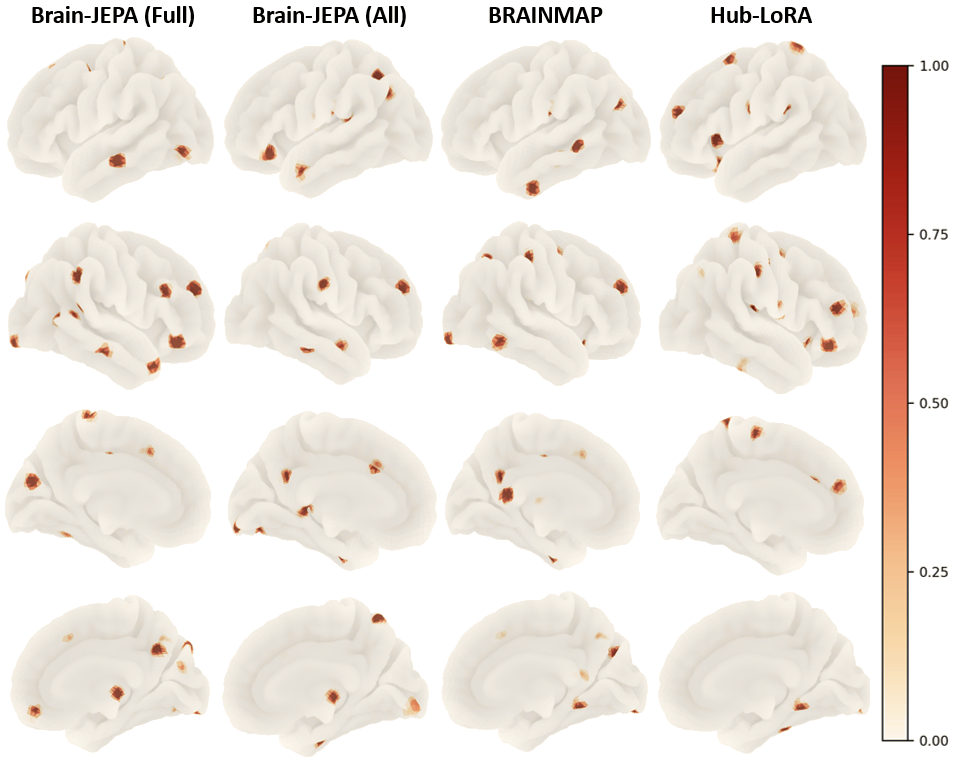}
   \caption{Surface plots based on normalized attributions produced via IG, for ADHD. Rows: Left lateral, Right lateral, Right medial, Left medial.}
   \label{fig:ig_adhd}
\end{figure}

Salient ADHD features (Figure \ref{fig:ig_adhd}\footnotemark[1]) exhibited a slightly different trend that follows the observation of BRAINMAP having lower Fidelity+ scores (on the ADHD dataset, as compared to ABIDE). 
BRAINMAP, in both settings (attention, IG), identified a wide-ranging set of features, including the insula, and DMN ROIs such as the posterior cingulate. However, it also introduces many ROIs with weak justification from the literature, such as uncus, declive and extra-nuclear. This could explain the low Fidelity+ scores, provide evidence of how having high model accuracy gives no guarantee that features identified from the model will be neurobiologically faithful, and demonstrate how RE-CONFIRM metrics can be used to reveal these insights. 
On the other hand, Brain-JEPA (IG) further includes ROIs from the frontal parietal network and thalamocortical control loops that are well associated with ADHD \cite{gao2019impairments}. Further finetuning with Hub-LoRA causes the model to focus on ROIs from the medial frontal gyrus, which are hubs known to be implicated in ADHD \cite{wang2025connections}. This also provides quantitative support for the observation of a significant increase in HAC.

For AD, we found that Brain-JEPA (All) highlights ROIs in the DMN (precuneus, inferior parietal lobe, middle frontal gyrus) and medial temporal lobe, which are well associated with AD. Following Hub-LoRA tuning, the model highlighted occipital regions such as inferior occipital gyrus and lingual gyrus, which are amongst the most disrupted brain hubs in AD \cite{gupta2020ambivert} and were not identified in the baseline model.

\subsubsection{Generalization to external unseen datasets}

\begin{table}[h]
\caption{\textcolor{black}{Out of distribution generalization for various approaches from Brain-JEPA (All) on ABIDE (columns 2-5) and ADHD-200 dataset (column 6). {\upshape\textsuperscript{\textasteriskcentered}} indicates p-value < 0.05}
\label{tab:ood}}
\centering
\resizebox{\columnwidth}{!}{
\begin{tabular}{l|llll|l}
\toprule
Metrics & BNI & EMC & GU & IP  & CNI-TLC \\
 & (29/29) & (27/27) & (51/55) & (20/20) & (100/100) \\
\midrule
With Hub-LoRA & \textbf{0.67\(\pm\)0.03} & \textbf{0.69\(\pm\)0.01} & \textbf{0.66\(\pm\)0.03} & \textbf{0.70\(\pm\)0.02} & \textbf{0.63\(\pm\)0.02} \\
No Hub-LoRA & 0.64\(\pm\)0.02 & 0.65\(\pm\)0.02\textsuperscript{\textasteriskcentered} & 0.62\(\pm\)0.01\textsuperscript{\textasteriskcentered} & 0.68\(\pm\)0.01 & 0.61\(\pm\)0.02 \\
Baseline (random) & 0.60\(\pm\)0.04\textsuperscript{\textasteriskcentered} & 0.63\(\pm\)0.03\textsuperscript{\textasteriskcentered} & 0.59\(\pm\)0.03\textsuperscript{\textasteriskcentered} & 0.61\(\pm\)0.02\textsuperscript{\textasteriskcentered} & 0.57\(\pm\)0.04\textsuperscript{\textasteriskcentered} \\
\bottomrule
\end{tabular}
}
\end{table}

\begin{figure}[h]
   \centering
   \includegraphics[width=0.95\linewidth]{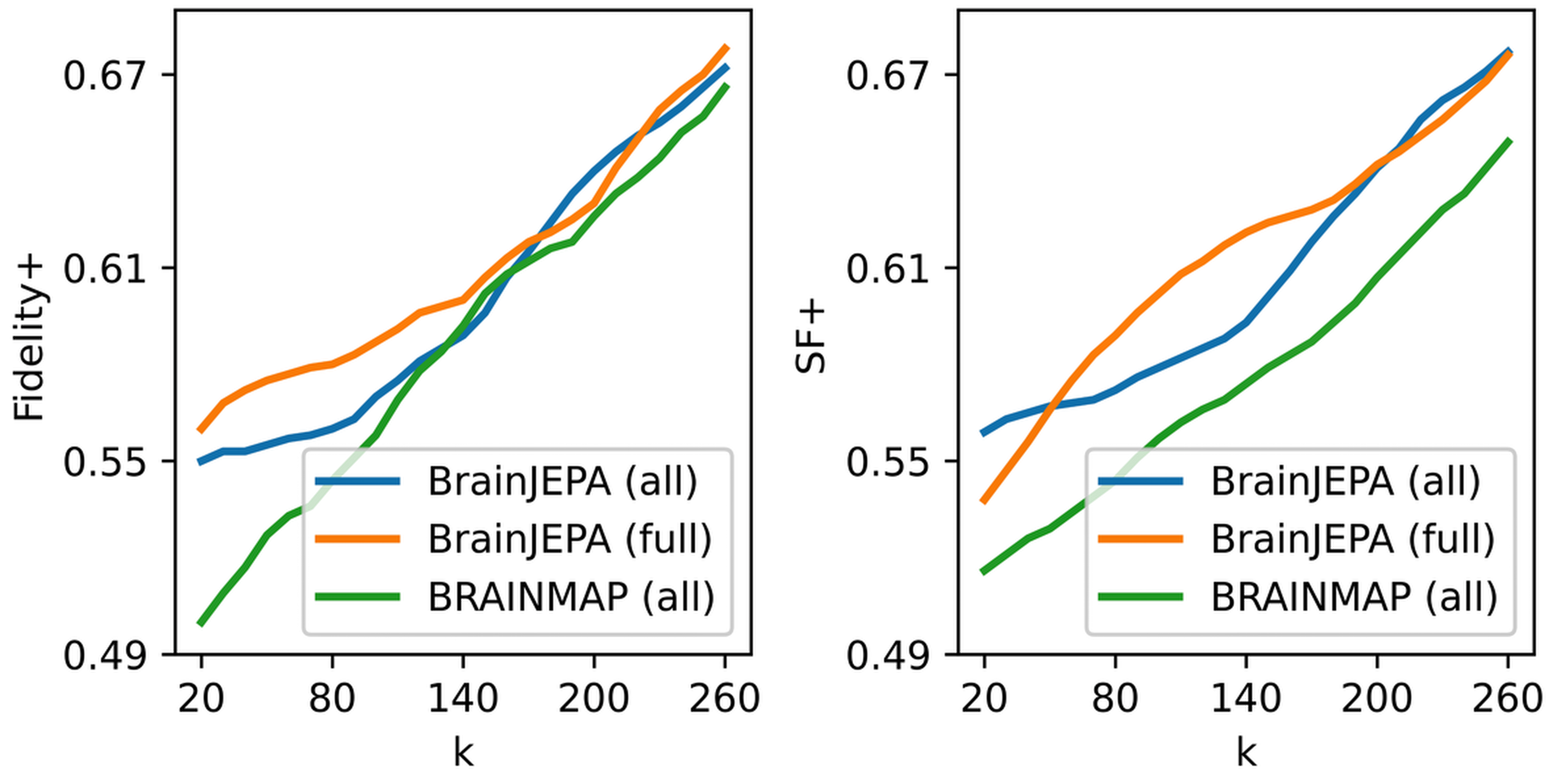}
   \caption{Evaluation of Fidelity+ and SF+ for different \textit{k} Features using IG attributor on the ABIDE dataset.}
   \label{fig:varykfidplus}
\end{figure}

\textcolor{black}{To assess the predictive utility of the potential biomarkers identified by RE-CONFIRM on unseen data, we evaluated their generalizability on the ABIDE-II dataset and CNI-TLC datasets.} Specifically, the top 20 features (20 was chosen based on the sensitivity analysis in Figure \ref{fig:varykfidplus}) identified by the Brain-JEPA (All) model after Hub-LoRA tuning were used to train an independent sigmoid-kernel support vector machine (SVM) on ABIDE-I and ADHD-200 datasets separately. The SVM hyperparameters were optimized using 5-fold cross-validation over the following ranges: $C \in \{0.1, 0.01, 0.001\}$, $\text{coef0} \in \{0.1, 0.01, 0.001\}$, and $\gamma \in \{10^{-1}, 10^{-3}, 10^{-5}\}$. The trained SVMs were then evaluated on four sites from the ABIDE-II dataset and the CNI-TLC dataset, respectively. We compared the results against randomly selected features and the top 20 features identified by other ablated settings, such as Brain-JEPA (All) without Hub-LoRA. \textcolor{black}{We used Welch's \(t\)-test and all comparisons showed large effect sizes Cohen's \(d \geq 1.0\).}

\textcolor{black}{Table \ref{tab:ood}\footnotemark[1] shows that features selected after Hub-LoRA tuning generalize better across out-of-distribution datasets, significantly outperforming random features at all sites and standard LoRA features at EMC and GU.} While randomly selected features could lead to a model performance slightly greater than chance, finetuning clearly helps to increase these improvements, and Hub-LoRA is more effective than just applying LoRA on all attention layers.

\subsubsection{Effects of varying metric-specific hyperparameters}

Values of Fidelity+ and SF+ are dependent on the threshold used ($k$). Thus, we varied this to study how the scores change. Our results confirmed the intuition that as more features are removed, the scores increase. However, the rate of increase varies across models and they seem to converge after about 50\% of features are removed. 
Thus, when fidelity metrics are used to compare across models, it has to be done at a fixed value of $k$ and smaller values of $k$ would be recommended as our results show that they are most sensitive when $k << 100$.

\section{Discussion}

One key takeaway from this study is the limitations of classification accuracy as a performance evaluation criterion. 
While this is useful to demonstrate that FMs are superior to baseline models (BrainGNN, STGCN, STAGIN) and comparable to specialised models (MSC-Mamba, BRAINMAP), our study reveals that it is inadequate for assessing biomarker robustness. 
Models with high classification performance do not necessarily translate to \textcolor{black}{potential biomarkers}. One possible reason is that the objective of minimising cross-entropy loss (empirical risk minimization) does not ensure that these models prioritise features that are neurobiologically faithful, nor does it guarantee that the model is not overfitted to the training set and robust to noise. 
Our proposed RE-CONFIRM framework addresses this by measuring five aspects of model robustness, encompassing both generic and biomarker-specific considerations. This goes beyond existing frameworks for evaluating XAI algorithms, such as 
Quantus \cite{hedstrom2023quantus}, as they are more generic and not designed for biomarker discovery. 

Our results reveal that FMs are superior in terms of classification accuracies as well as non-biomarker specific metrics of robustness - FMs have better IS, LRC and MPRC as compared to specialised, non-FM models like BRAINMAP. However, we also found that much care has to be taken in the process of finetuning FMs. While finetuned FMs are clearly and intuitively superior to linear probing (i.e. only finetuning the classifier), full finetuning on FC datasets does not always lead to better results than more constrained forms of finetuning, such as LoRA, in the context of dFC data. This does not necessarily suggest that LoRA always outperforms full finetuning - rather, it highlights how existing FM finetuning techniques seem to depend on ad-hoc, empirical adjustments for it to work on connectome datasets. Similar observations have been made in Electroencephalography (EEG) research, where they found that constrained finetuning outperformed full tuning \cite{cui2024neuro}.
Our results further showed that this applies to FMs of various pre-training methodologies, such as masked signal modelling or autoregression. These consistent observations could possibly be explained by how connectome datasets often have much smaller dataset sizes relative to their high dimensionality, making it challenging to use such small data to tune the millions of parameters used in FMs. Also, we found that simple finetuning of all layers fails to capture salient features that are well-supported by existing literature. Our experiments give greater clarity as to how FMs should be finetuned and our proposed Hub-LoRA technique makes it possible for FMs to match the performance of other \ac{DL} models that are specialized to capture connectome-specific properties, such as the involvement of hubs in neurological disorders.

Finally, as seen in experiments, the choice of XAI algorithms has a very significant role in influencing the salient features identified from DL models. Many papers report potential biomarkers, but fail to consider how they vary when a different feature attributor is used, even when the predictor is kept the same.
Our previous work identified that the use of GNNExplainer is superior to attention-based approaches and weight pooling \cite{girish2024robustness}. In this work, we performed a deeper analysis on more generally applicable attributors (IG and attention), revealing that for ASD, IG produced more biologically plausible biomarkers (default mode network, salience network, thalamocortical and fusiform areas \cite{guo2024systematic}) while attention scores over-emphasise ROIs related to sensory and visual cortex that are less well backed by existing literature and meta-analyses. 
\textcolor{black}{This suggests that future studies proposing novel predictors should evaluate them with multiple attributors, identify the best-performing combinations, and assess their neurobiological plausibility before reporting salient features.}

One limitation of our study is that the scope is limited to brain graphs. We have not explored the robustness of biomarkers derived from population graphs. Population graphs allow a wide variety of multimodal data to be integrated in the analysis via the construction of the graph. While factors such as the choice of GNN and graph construction (e.g. node homophily) have been identified to be key drivers of performance \cite{muller2024population}, their impact on model interpretability has yet to be examined thoroughly. 
Additionally, future work could investigate how theoretical insights into \ac{XAI} algorithms \cite{huang2024failings} can further guide the selection of attributors, beyond the empirical approach adopted in this study.

Overall, brain FMs are superior to other DL models in terms of classification performance, but biomarker discovery demands a stricter set of requirements beyond such traditional metrics. 
Task-dependent robustness specifications \cite{xian2025robustness} have been recently suggested as an approach to assess the robustness of biomedical FMs. Seen in this context, RE-CONFIRM metrics serve as a form of domain-specific tests to ensure that attributions generated from FMs are reasonable, and Hub-LoRA provides a way to improve sensitivity to hubs in FMs. Our study lays the groundwork towards developing a set of robustness specifications for ML-driven biomarker discovery.

\section{Conclusion}

\textcolor{black}{In conclusion, we demonstrated that models with the highest accuracies do not always lead to neurobiologically plausible biomarkers.} Our proposed RE-CONFIRM framework, comprising five evaluation metrics, provides a more comprehensive view of model performance by augmenting them with existing performance measures. 
Our results support the use of FMs for biomarker discovery as it performs well in terms of classification performance and generic metrics such as stability. However, bFMs are found to be less sensitive to hubs, and simply finetuning all their layers does not improve this. \textcolor{black}{Thus, we proposed Hub-LoRA as a way to finetune bFMs to improve their hub sensitivity, yielding potential biomarkers that are more neurobiologically plausible.} 
Finally, we provided additional evidence that demonstrates how simply choosing different attributors results in vastly different salient features being reported. Future studies that propose architectural improvements for disorder prediction tasks should experiment with various attributors before reporting salient features. RE-CONFIRM could be used to evaluate biomarker robustness and guide the choice of the final predictor-attributor combination. \textcolor{black}{Adoption of more robust evaluation approaches beyond model accuracy could be key for catalysing progress towards discovering reproducible biomarkers.}

\section*{References}

\bibliographystyle{ieeetr}
\bibliography{reference}

\section{Appendix}

\setcounter{table}{0}
\renewcommand{\thetable}{S\arabic{table}}
\setcounter{figure}{0}
\renewcommand{\thefigure}{S\arabic{figure}}

\begin{table}[!ht]
\caption{Additional comparison of classification accuracies (\%) for different number of trainable model parameters. \label{tab:accuracies-additional}}
\centering
\resizebox{\columnwidth}{!}{
\begin{tabular}{llllll}
\toprule
Models & Type & ABIDE & ADHD & ADNI & Params \\
\midrule
BrainGNN-1M & sFC & 62.20 \(\pm\) 3.65 & 60.11 \(\pm\) 3.47 & 60.32 \(\pm\) 4.11 & 1,000k \\
\midrule
STGCN-1M & dFC & 61.23 \(\pm\) 4.07 & 62.46 \(\pm\) 4.34 & 62.60 \(\pm\) 4.73 & 1,000k \\ 
BRAINMAP-13M & dFC & 68.52 \(\pm\) 3.92 & 69.94 \(\pm\) 4.78 & 70.15 \(\pm\) 4.12 & 13,320k \\
\midrule
BrainLM-111M & bFM & 72.66 \(\pm\) 3.18  &  71.20 \(\pm\) 2.96 & 73.48 \(\pm\) 3.55 & 111,460k \\
\bottomrule
\end{tabular}}
\end{table}

\begin{table}[ht]
\caption{RE-CONFIRM evaluation metrics for attention attributor.
\label{tab:metrics-attention}}
\centering
\resizebox{\columnwidth}{!}{
\begin{tabular}{l|lll|lll}
\toprule
 & \multicolumn{3}{c|}{ABIDE} & \multicolumn{3}{c}{ADHD-200} \\
\cmidrule(lr){2-4} \cmidrule(lr){5-7}
Metrics & BrainLM & Brain-JEPA & BRAINMAP & BrainLM & Brain-JEPA & BRAINMAP \\
\midrule
IS ($\downarrow$) & \textbf{5.16\(\pm\)1.08} & 6.03\(\pm\)0.76 & 7.72\(\pm\)1.34 & 5.95\(\pm\)0.87 & \textbf{5.32\(\pm\)1.41} & 7.89\(\pm\)1.64 \\
LRC ($\uparrow$) & 0.49\(\pm\)0.03 & \textbf{0.53\(\pm\)0.03} & 0.51\(\pm\)0.01 & 0.54\(\pm\)0.03 & \textbf{0.56\(\pm\)0.02} & 0.49\(\pm\)0.02\\
MPRC ($\uparrow$) & \textbf{0.56\(\pm\)0.02} & 0.55\(\pm\)0.02 & 0.54\(\pm\)0.02 & 0.55\(\pm\)0.03 & \textbf{0.60\(\pm\)0.02} & 0.59\(\pm\)0.03\\
\midrule
SF+ ($\uparrow$) & 0.49\(\pm\)0.03 & \textbf{0.55\(\pm\)0.02} & 0.52\(\pm\)0.01 & 0.49\(\pm\)0.02 & \textbf{0.55\(\pm\)0.03} & 0.53\(\pm\)0.02 \\
HSI ($\uparrow$) & 0.69\(\pm\)0.02 & 0.71\(\pm\)0.01 & \textbf{0.75\(\pm\)0.02} & 0.69\(\pm\)0.03 & 0.72\(\pm\)0.02 & \textbf{0.76\(\pm\)0.02}\\
\midrule
Fidelity+ ($\uparrow$) & 0.55\(\pm\)0.02 & \textbf{0.57\(\pm\)0.02} & 0.51\(\pm\)0.02 & \textbf{0.53\(\pm\)0.02} & 0.52\(\pm\)0.03 & 0.39\(\pm\)0.03\\
HAC ($\uparrow$) & 0.26\(\pm\)0.03 & 0.27\(\pm\)0.02 & \textbf{0.34\(\pm\)0.03} & 0.25\(\pm\)0.02 & 0.23\(\pm\)0.02 & \textbf{0.37\(\pm\)0.02} \\
\bottomrule
\end{tabular}
}
\end{table}

\begin{figure}[h]
   \centering
   \includegraphics[width=0.95\linewidth]{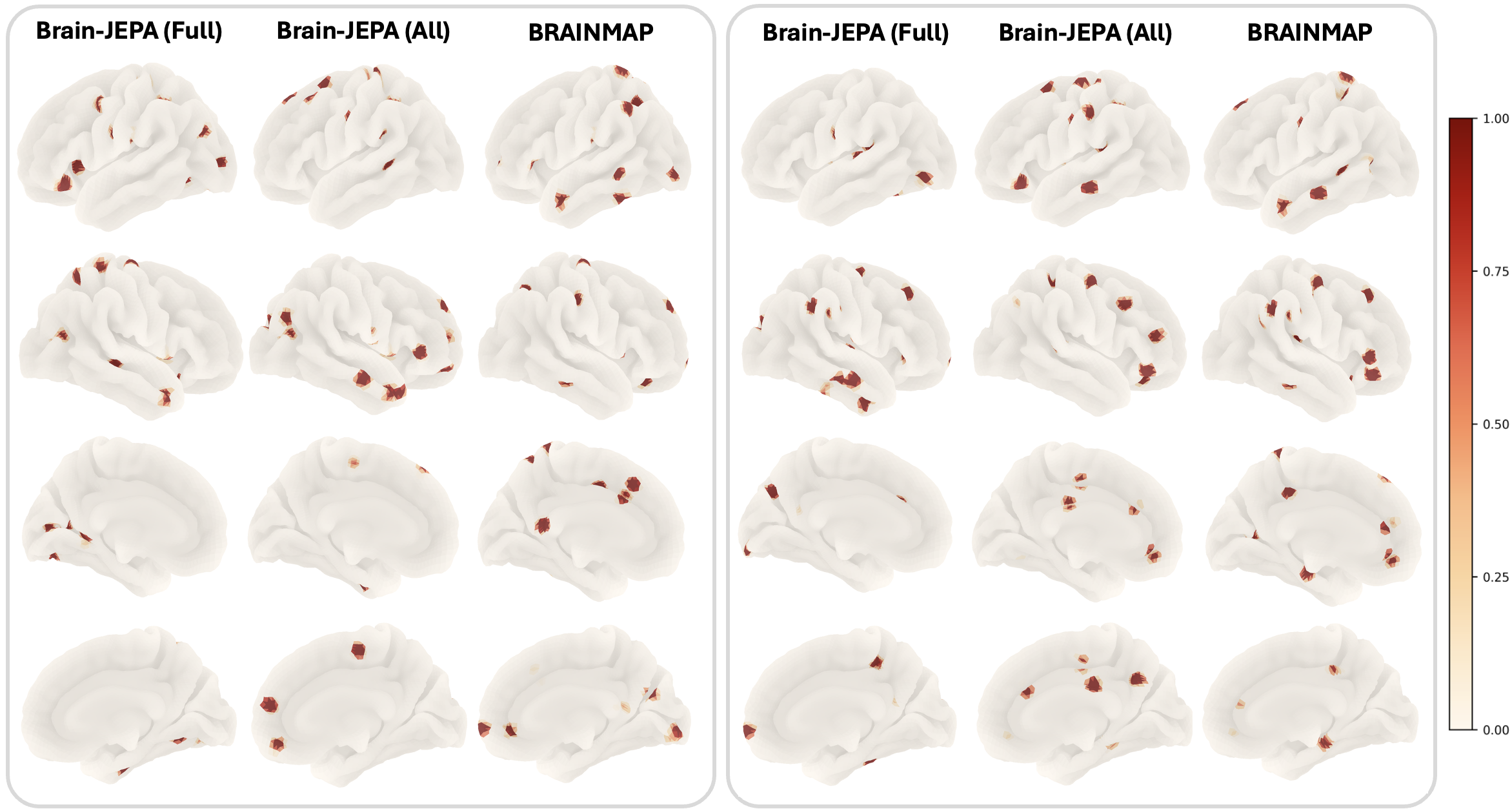}
   \caption{Surface plots based on normalized attribution scores produced by various models from attention, for ADHD (left columns) and ABIDE dataset (right columns). Rows: Left lateral, Right lateral, Right medial, Left medial.}
   \label{fig:attn}
\end{figure}

\end{document}